# Impedances of laminated vacuum chambers

*A. Burov and V. Lebedev, Fermilab*

*Abstract*

Longitudinal and transverse impedances are derived for round and flat laminated vacuum chambers.

## *Introduction*

First publications on impedance of laminated vacuum chambers are related to early 70-s: those are of S. C. Snowdon [1] and of A. G. Ruggiero [2]; fifteen year later, a revision paper of R. Gluckstern appeared [3]. All the publications were presented as Fermilab preprints, and there is no surprise in that: the Fermilab Booster has its laminated magnets open to the beam. Being in a reasonable mutual agreement, these publications were all devoted to the longitudinal impedance of round vacuum chambers. The transverse impedance and the flat geometry case were addressed in more recent paper of K. Y. Ng [4]. The latest calculations of A. Macridin et al. [5] revealed some disagreement with Ref. [4]; this fact stimulated us to get our own results on that matter. As it can be seen below, results of this paper agree with Ref. [5].

Some general conditions are assumed here. First, the frequencies under interest, $\omega$, are supposed to be sufficiently low [6.7]:

$$\omega \ll \gamma\beta c / a;$$
$$\omega \ll 4\pi\sigma / \varepsilon.$$

Here $a$ is the aperture radius, $\gamma$ and $\beta$ are the relativistic factors, $c$ is the speed of light, $\sigma$ and $\varepsilon$ are the chamber conductivity and dielectric constant. The first condition actually requires the wavelength of the fields to be much longer than the aperture, as they are seen in the beam frame. Note that the specified wavelength parameter $\gamma\beta c / (a\omega)$ is relevant to the wake forces, not to the electric and magnetic fields taken separately. For the separate field components, the relativistic factor does not count; but it does count for the wakes (see e. g. Ref. [8], Eq. (2.41))

The above condition seems to be satisfied for all practically interesting cases. It allows one to neglect the longitudinal magnetic field, and, consequently, the transverse components of the vector potential vanish. The second condition means that the beam electric moments are shielded infinitesimally fast at the chamber surface. While this condition is well satisfied for metals, it may be violated for ferrites [9]. The last case is irrelevant to this paper, since the laminations are metallic (iron). We also imply that the laminations are thin: $h, d \ll a$, and that the skin depth, $\delta$, is much smaller than the lamination thickness, $d$.

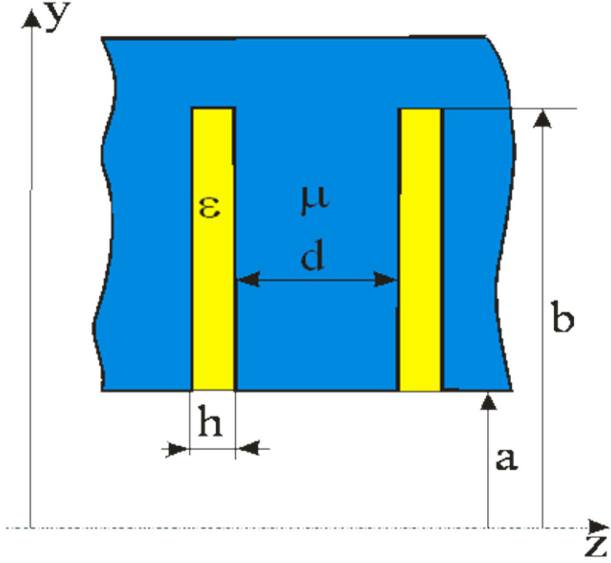

Figure 1. Geometry of laminated vacuum chamber.

## 1. Flat chamber

### Longitudinal impedance

Let the beam current be modulated at a frequency $\omega$:

$$I(\mathbf{r},t) = I_0 \delta(\mathbf{r}_\perp) \exp(-i\omega(t - z/\mathrm{v})). \tag{1.1}$$

Due to the horizontal homogeneity, the problem can be solved by the Fourier-transform over this coordinate:

$$F(x) = \int_{-\infty}^{\infty} F_k \exp(ik_x x) \frac{dk_x}{2\pi}. \tag{1.2}$$

Since only the Fourier components are used below, the subscript $k$ can be safely omitted. For long wavelength, the vector potential reduces to its longitudinal component only. In the free

space, it satisfies the transverse Laplace equation and can be presented as

$$A = \frac{I_0 Z_0}{2k_x}\left[\exp(-k_x y) - G\frac{\cosh(k_x y)}{\cosh(k_x a)}\exp(-k_x a)\right]; \quad k_x > 0, \ 0 < y < a, \quad (1.3)$$

where $Z_0 = 4\pi/c = 377$ Ohm, $a$ is the half-gap (see Fig. 1); $G = G(k_x)$ is the function to be determined from the boundary conditions, and the vector potential is an even function of $k_x$ and $y$. The first term inside the square brackets describes a direct field of the beam, while the second one is the response due to the induced currents. From here, a ratio of the magnetic fields follows:

$$\left.\frac{H_y}{H_x}\right|_{y=a-0} = i\frac{1-G}{1+G\tanh(k_x a)}. \quad (1.4)$$

Using the boundary conditions at the metal surface one can easily prove that the vector potential inside a thin crack satisfies the Helmholtz equation:

$$\Delta_\perp A^{crack} = -k^2 A^{crack}, \quad (1.5)$$

where

$$k^2 \equiv \frac{\omega^2 \varepsilon}{c^2}\left(1 + \frac{2\mu}{\kappa h}\right) \equiv \frac{\omega^2 \varepsilon}{c^2} + g^2; \quad \kappa \equiv \frac{1-i\,\text{sgn}\,\omega}{\delta} \equiv (1-i\,\text{sgn}\,\omega)\frac{\sqrt{2\pi|\omega|\sigma\mu}}{c}. \quad (1.6)$$

Note that Eq. (1.6) is only justified if $gh/2 \ll 1$. In that case the fields inside the crack can be treated as independent from the $z$-coordinate (coordinate normal to its surface). Otherwise one need to take into account that the fields in the crack are dependent on $z$ as $\cosh(gz)$ or $\sinh(gz)$, resulting in a more complicated form for Eq. (1.6). In most practical cases the thin crack approximation is valid. Taking into account that the crack is shorted at $y = b$, the fields can be written inside the crack as:

$$\begin{aligned}
A^{crack} &= A_0 \sin(k_y(b-y)); \quad k_y = \sqrt{k^2 - k_x^2}; \\
H_x^{crack} &= -k_y A_0 \cos(k_y(b-y)); \\
H_y^{crack} &= -ik_x A_0 \sin(k_y(b-y)).
\end{aligned} \quad (1.7)$$

A vertical magnetic flux through the metal surface is

$$\left.\int B_y dz\right|_{metal} = -2ik_x \mu A_0 \sin(k_y(b-y))/\kappa, \quad (1.8)$$

where the factor of 2 comes out due to the two sides of the lamina. Adding the flux through the crack itself, one obtains the average magnetic field, is

$$\bar{B}_y \equiv \frac{1}{d+h}\left(\int B_y dz\Big|_{crack} + \int B_y dz\Big|_{metal}\right) = -ik_x A_0\left(1+\frac{2\mu}{\kappa h}\right)\frac{h}{d+h}\sin(k_y(b-y)), \quad (1.9)$$

yielding

$$\frac{\bar{B}_y}{H_x^{crack}}\bigg|_{y=a+0} = \frac{ik_x}{k_y}\frac{h}{d+h}\left(1+\frac{2\mu}{\kappa h}\right)\tan\left(k_y(b-a)\right) \equiv iR_B \quad (1.10)$$

The condition $y = a+0$ means staying vertically at $y = a+\Delta y$ so that $h, \delta \ll \Delta y \ll 1/k$. Similarly, $y = a-0$ means $y = a-\Delta y$. Since both the average magnetic field, Eq. (1.9), and the horizontal field at the crack region are preserved at crossing the magnet border $y=a$, their ratio is preserved as well:

$$\frac{\bar{B}_y}{H_x^{crack}}\bigg|_{y=a+0} = \frac{H_y}{H_x}\bigg|_{y=a-0} \quad (1.11)$$

Thus, Eqs. (1.10) and (1.4) lead to the induced field amplitude

$$G = \frac{1-R_B}{1+R_B \tanh(k_x a)} \quad (1.12)$$

At this point, only an average electric field has to be found. To do that, the Maxwell equation

$$-\frac{\partial E_z}{\partial x} + \frac{\partial E_x}{\partial z} = i\frac{\omega}{c}B_y \quad (1.13)$$

can be averaged over a period, yielding

$$\bar{E}_z\Big|_{y=a+0} = -\frac{\omega}{ck_x}\bar{B}_y\Big|_{y=a+0} \quad (1.14)$$

Average electric fields above and below the boundary (with a thickness of $\Delta y$) are related as

$$\bar{E}_z\Big|_{y=a+0} - \bar{E}_z\Big|_{y=a-0} = i\frac{\omega\mu}{\kappa c}H_x\Big|_{y=a-0} \quad (1.15)$$

Using Eq. (1.12), the horizontal field is found:

$$H_x\Big|_{y=a-0} = \frac{\partial A}{\partial y} = -\frac{Z_0 I_0}{2\cosh(k_x a)}\frac{1}{1+R_B \tanh(k_x a)}. \quad (1.16)$$

Finally Eq. (1.16), (1.14) and (1.10) yield the following result,

$$Z_\| = -\frac{1}{I_0}\int_0^\infty \bar{E}_z\Big|_{y=0}\frac{dk_x}{\pi} = -i\frac{\omega}{c}\frac{Z_0}{2\pi}\int_0^\infty \frac{dk_x}{k_x}\frac{R_B + \mu k_x/\kappa}{\cosh^2(k_x a)\left(1+R_B \tanh(k_x a)\right)}, \quad (1.17)$$

for the longitudinal impedance per unit length follows. Here we used that

$$\left.\bar{E}_z\right|_{y=0} = \frac{\left.\bar{E}_z\right|_{y=a-0}}{\cosh(k_x a)} \ . \tag{1.18}$$

*Transverse impedance*

For the horizontal beam oscillations, the vector potential is an even function of the vertical coordinate and odd one of the horizontal; according to [6]

$$A = -i\frac{D_0 Z_0}{2}\left[\exp(-k_x y) - G\frac{\cosh(k_x y)}{\cosh(k_x a)}\exp(-k_x a)\right]; \ k_x > 0, \ 0 < y < a, \tag{1.19}$$

with $D_0$ as the amplitude of the beam dipole moment oscillations. Note that this field differs from the longitudinal case, Eq. (1.3), only by the amplitude; thus, all the field ratios remain the same. In particular, Eq. (1.12) is valid for this case as well. Using Eqs. (10, 11) of Ref. [6], the horizontal impedance follows:

$$Z_x = Z_x^\sigma + Z_x^\infty = -i\frac{Z_0 \beta}{2\pi}\int_0^\infty \frac{(R_B + \mu k_x/\kappa)k_x dk_x}{\cosh^2(k_x a)(1 + R_B \tanh(k_x a))} - i\frac{Z_0}{2\pi a^2 \beta\gamma^2}\frac{\pi^2}{24} \ . \tag{1.20}$$

The vertical impedance can be found from the horizontal by a substitution $\cosh(k_x a) \leftrightarrow \sinh(k_x a)$ in the finite conductance term $Z_x^\sigma$, and taking twice higher infinite conductivity term [6]:

$$Z_y = Z_y^\sigma + Z_y^\infty = -i\frac{Z_0 \beta}{2\pi}\int_0^\infty \frac{(R_B + \mu k_x/\kappa)k_x dk_x}{\sinh^2(k_x a)(1 + R_B \coth(k_x a))} - i\frac{Z_0}{2\pi a^2 \beta\gamma^2}\frac{\pi^2}{12} \ . \tag{1.21}$$

Note that the second terms in the integrand numerator in Eqs. (1.17), (1.20) and (1.21) ($\mu k_x/\kappa$) yields the conventional resistive wall impedances when the crack width approaches zero.

## 2. Round Chamber

*Longitudinal impedance*

For a round vacuum chamber of radius $a$ and arbitrary walls, the axially symmetric fields in the free space are related so that (Ref. [8], Eq. (2.3)) :

$$H_\varphi = \frac{2I_0}{rc} - i\frac{\omega r}{2c}E_z, \ r < a . \tag{2.1}$$

From here, the longitudinal impedance $Z_\parallel$ can be related to the so-called surface impedance $R$:

$$Z_\parallel = -\frac{E_z}{I_0} = \frac{Z_0}{2\pi a} \frac{R}{1 - i\omega a R/(2c)} \ ; \quad R \equiv -E_z/H_\varphi\big|_{r=a-0}. \tag{2.2}$$

The Maxwell Equation $\nabla \times \mathbf{E} = i\omega \mathbf{B}/c$, applied to the azimuthal direction, relates inner and outer average longitudinal electric fields (compare with Eq. (1.15)):

$$\bar{E}_z\big|_{r=a+0} - \bar{E}_z\big|_{r=a-0} = -i\omega \mu H_\varphi/(\kappa c). \tag{2.3}$$

This can also be written as

$$R = R_+ - i\frac{\omega\mu}{\kappa c}; \quad R_+ \equiv -\frac{\bar{E}_z\big|_{r=a+0}}{H_\varphi} \tag{2.4}$$

Inside the crack, the longitudinal electric field satisfies the Helmholtz equation (compare with Eq. (1.5)):

$$\Delta_\perp E_z^{\text{crack}} = -k^2 E_z^{\text{crack}} \ ;$$
$$H_\varphi^{\text{crack}} = i\frac{\omega\varepsilon}{k} \frac{\partial E_z^{\text{crack}}}{\partial(kr)}. \tag{2.5}$$

From here, the field components are expressed in terms of the Hankel functions:

$$E_z^{\text{crack}} = E_0 \left[ H_0^{(1)}(kr)H_0^{(2)}(kb) - H_0^{(2)}(kr)H_0^{(1)}(kb) \right];$$
$$H_\varphi^{\text{crack}} = -i\frac{\omega\varepsilon}{ck} E_0 \left[ H_1^{(1)}(kr)H_0^{(2)}(kb) - H_1^{(2)}(kr)H_0^{(1)}(kb) \right]. \tag{2.6}$$

A factor $\cosh(gz)$ is omitted according to the assumption $gh/2 \ll 1$. Since there is no longitudinal electric field in the metal, only the crack electric field contributes to its average:

$$\bar{E}_z\big|_{r=a+0} = E_z^{\text{crack}} \frac{h}{d+h}. \tag{2.7}$$

Together with Eq. (2.6), this yields

$$R_+ \equiv -\frac{\bar{E}_z}{H_\varphi^{\text{crack}}}\bigg|_{r=a+0} = -i\frac{ckh}{\omega\varepsilon(d+h)} \frac{H_0^{(1)}(ka)H_0^{(2)}(kb) - H_0^{(2)}(ka)H_0^{(1)}(kb)}{H_1^{(1)}(ka)H_0^{(2)}(kb) - H_1^{(2)}(ka)H_0^{(1)}(kb)}. \tag{2.8}$$

With Eq. (2.4), the impedance (2.2) follows:

$$Z_\parallel = \frac{Z_0}{2\pi a} \frac{R_+ - i\omega\mu/(\kappa c)}{1 - i\frac{\omega a R_+}{2c} - \frac{\omega^2 a\mu}{2\kappa c^2}}, \tag{2.9}$$

where the second term in the numerator is responsible for the conventional resistive wall impedance when the cracks disappear.

*Transverse impedance*

For the transverse dipole oscillations, the vector potential in the free space can be written as

$$A = \frac{2D_0}{ca}\left(\frac{a}{r} - G\frac{r}{a}\right)\cos\varphi \equiv A_0\left(\frac{a}{r} - G\frac{r}{a}\right)\cos\varphi, \tag{2.10}$$

where $D_0$ is the amplitude of the dipole moment oscillations. In terms of the induced field amplitude $G$, the transverse impedance is expressed as [7]

$$Z_\perp = Z_\perp^\sigma + Z_\perp^\infty = -i\frac{Z_0\beta(1-G)}{2\pi a^2} - i\frac{Z_0}{2\pi a^2\beta\gamma^2}. \tag{2.11}$$

At the inner border, $r = a - 0$, the longitudinal electric and azimuthal magnetic fields follow as

$$\begin{aligned}E_z &= i\omega A/c = i\omega A_0(1-G)\cos\varphi/c; \\ H_\varphi &= -\partial A/\partial r = A_0(1+G)\cos\varphi/a.\end{aligned} \tag{2.12}$$

This relates the surface impedance $R = -E_z/H_\varphi\big|_{r=a-0}$ and the induced field amplitude $G$:

$$R = -i\frac{\omega a}{c}\frac{1-G}{1+G} \Leftrightarrow 1-G = \frac{2R}{R - i\omega a/c} \tag{2.13}$$

Note that although the fields $E_z, H_\varphi$, etc. and their ratios $R, R_+$ are denoted by the same symbols for the longitudinal and the transverse cases, they are not the same and should not be confused. Inside the crack, the field components $E_z^{crack}, H_\varphi^{crack}$ satisfy Eq. (2.5), leading for the dipole mode to

$$\begin{aligned}E_z^{crack} &= E_0\left[H_1^{(1)}(kr)H_1^{(2)}(kb) - H_1^{(2)}(kr)H_1^{(1)}(kb)\right]\cos\varphi; \\ H_\varphi^{crack} &= i\frac{\omega\varepsilon}{ck}E_0\left[H_1^{(1)'}(kr)H_1^{(2)}(kb) - H_1^{(2)'}(kr)H_1^{(1)}(kb)\right]\cos\varphi.\end{aligned} \tag{2.14}$$

For the calculations, it is useful to remember the derivatives of the Hankel functions are expressed as

$$H_1'(x) = [H_0(x) - H_2(x)]/2. \tag{2.15}$$

Equations (2.14) yield the field ratio

$$R_+ \equiv -\frac{\overline{E}_z}{H_\varphi^{crack}}\bigg|_{r=a+0} = i\frac{ckh}{\omega\varepsilon(d+h)}\frac{H_1^{(1)}(ka)H_1^{(2)}(kb) - H_1^{(2)}(ka)H_1^{(1)}(kb)}{H_1^{(1)'}(ka)H_1^{(2)}(kb) - H_1^{(2)'}(ka)H_1^{(1)}(kb)} \tag{2.16}$$

With Eqs. (2.4) and (2.13), this formula yields the transverse impedance (2.11)

$$Z_\perp = Z_\perp^\sigma + Z_\perp^\infty = -i\frac{Z_0 \beta}{\pi a^2}\frac{R}{R - i\omega a/c} - i\frac{Z_0}{2\pi a^2 \beta \gamma^2};$$

$$R = R_+ - i\frac{\omega\mu}{\kappa c} = i\frac{ckh}{\omega\varepsilon(d+h)}\frac{H_1^{(1)}(ka)H_1^{(2)}(kb) - H_1^{(2)}(ka)H_1^{(1)}(kb)}{H_1^{(1)\prime}(ka)H_1^{(2)}(kb) - H_1^{(2)\prime}(ka)H_1^{(1)}(kb)} - i\frac{\omega\mu}{\kappa c}.$$

(2.17)

## *3. Discussion*

It would be good to discuss the impedances on a base of real parameters of the Booster magnets. However, some of the important parameters are actually unknown. While the inner and outer aperture *a* and *b*, as well as the lamina thickness *d* are perfectly known, we have a poor knowledge of the magnetic permeability $\mu$ at the interesting frequency range of hundreds MHz. Moreover, the guiding magnetic field makes that value not just a function of frequency, but a tensor function. Another uncertainty relates to the crack width *h*. Comparison of the average lamina thickness with the entire length of the magnet gives only a magnet-average value for *h*. There is no reason to assume that these values have a narrow distribution over their average, but the rms of that distribution is unknown. Ideally, the calculated impedances have to be averaged over this distribution – but it cannot be done even approximately without knowing the rms spread of the crack widths. One more uncertainty relates to thickness of the iron oxide at the lamina surfaces, which may change the crack properties. All these uncertainties can be reduced with a set of dedicated measurements, which hopefully will be made in future. Some measurements were already performed [10], and they allow to see how good or bad is one or another hypothesis about the unknown values. Without any strong statement about our choice, we are assuming parameters of Table 1 for the Booster F-magnet, which show a reasonably good agreement with the measurements of Ref. [10]. With new measurements, the values of some parameters should be known better.

| $a$ | Magnet half-gap | 2.08 cm |
|---|---|---|
| $b$ | Outer short-cut | 16.5 cm |
| $d$ | Lamina thickness | 0.064 cm |
| $h$ | Crack width | 0.002 cm |
| $\sigma$ | Conductivity | $4.5 \cdot 10^{16}$ 1/s |
| $\varepsilon$ | Dielectric permittivity | 4.75 |
| $\mu$ | Magnetic permeability | 50 |

Table 1: accepted parameters of the Booster F-magnet.

Longitudinal impedances for that magnet, Eq. (1.17), (2.9) are shown in Fig. 2.

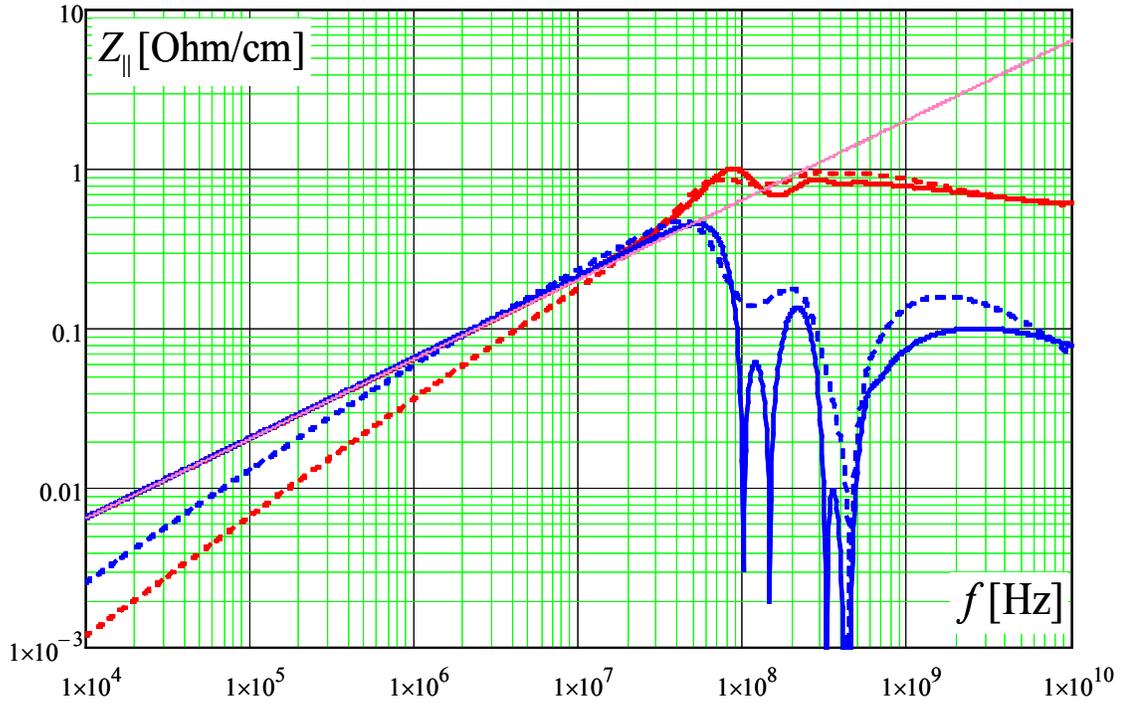

Figure 2: Longitudinal impedances for the round (solid lines) and flat (dash lines) geometries. Red lines are for the real parts, blue – for the absolute value of the imaginary parts. The magenta line shows the low-frequency approximation $\operatorname{Re} Z_{\|}^{LF} = \dfrac{1}{\pi d \sigma \delta} \ln(b/a) = \operatorname{Re} Z_{\|}^{conv} \dfrac{2a}{d} \ln(b/a)$ with $Z_{\|}^{conv} = \dfrac{\kappa}{2\pi a \sigma}$ as longitudinal impedance of the conventional solid round vacuum chamber of the same metal.

Several features of Fig. 2 deserve to be noted.

- The low limit of the frequency range is determined by the skin depth: at 10 kHz $\delta \approx d/2$.
- At low frequencies, $f ≲ 50$MHz, a simplistic electrotechnical approximation $Z_\parallel = \frac{\kappa}{\pi d \sigma}\ln(b/a) = Z_\parallel^{conv}\frac{2a}{d}\ln(b/a) \propto \omega^{1/2}$ for the round geometry coincides with the actual solution. For the flat case, the low-frequency impedance behavior is different, $Z_\parallel \propto \omega^{3/4}$.
- Note that impedance of the conventional solid vacuum chamber $Z_\parallel^{conv} = \frac{\kappa}{2\pi a \sigma}$ exceeds the careless limit $|Z_\parallel/n| \leq Z_0/2$ [8] by a factor of $(\mu\delta/a)\ln(b/a)$. For $\mu \gg 1$ this can be a big number. The reason is that the field energy located inside the magnetic chamber grows unlimitedly with the magnetic permeability: $\frac{\mu H^2}{8\pi}2\pi a \delta \propto \sqrt{\mu}$.
- A limit for the low-frequency approximation is determined by the field decay along the crack depth, $\operatorname{Im} k \propto \omega^{3/4}$, see Eq. (1.6). At sufficiently high frequency, when $\operatorname{Im} kb \gg 1$, this radial field decay limits the length of the shielding current along the crack surface before it reaches the outer short-cut radius $b$. At $f ≳ 1$ GHz, $\operatorname{Im} ka \geq 1$, so the path length of the shielding current gets proportional to the field decay length $\operatorname{Im} k$, leading to $Z_\parallel \simeq Z_\parallel^{conv}\frac{2}{d\operatorname{Im} k} \propto \omega^{-1/4}$
- For the conventional solid vacuum chambers, the longitudinal emittance of the flat chamber is known to be equal to one of the round chamber [11,12]. In other words, the longitudinal Yokoya factor of the solid flat chamber, or the ratio of flat-to-round impedances is 1. As it is seen from Fig. 2, the Yokoya factor of the flat laminated chamber is close to 1 at $f \geq 10$ MHz, while at lower frequencies it may be significantly smaller.

The transverse impedances are presented in Fig. 3. There are several reasons for the complicated behavior of the transverse impedances. First, the depth of field penetration inside the crack changes at $|ka| \sim 1$. Above that frequency (~ 1GHz), the shielding current path length is determined by the decay along the crack, while below that it is determined by the aperture $a$.

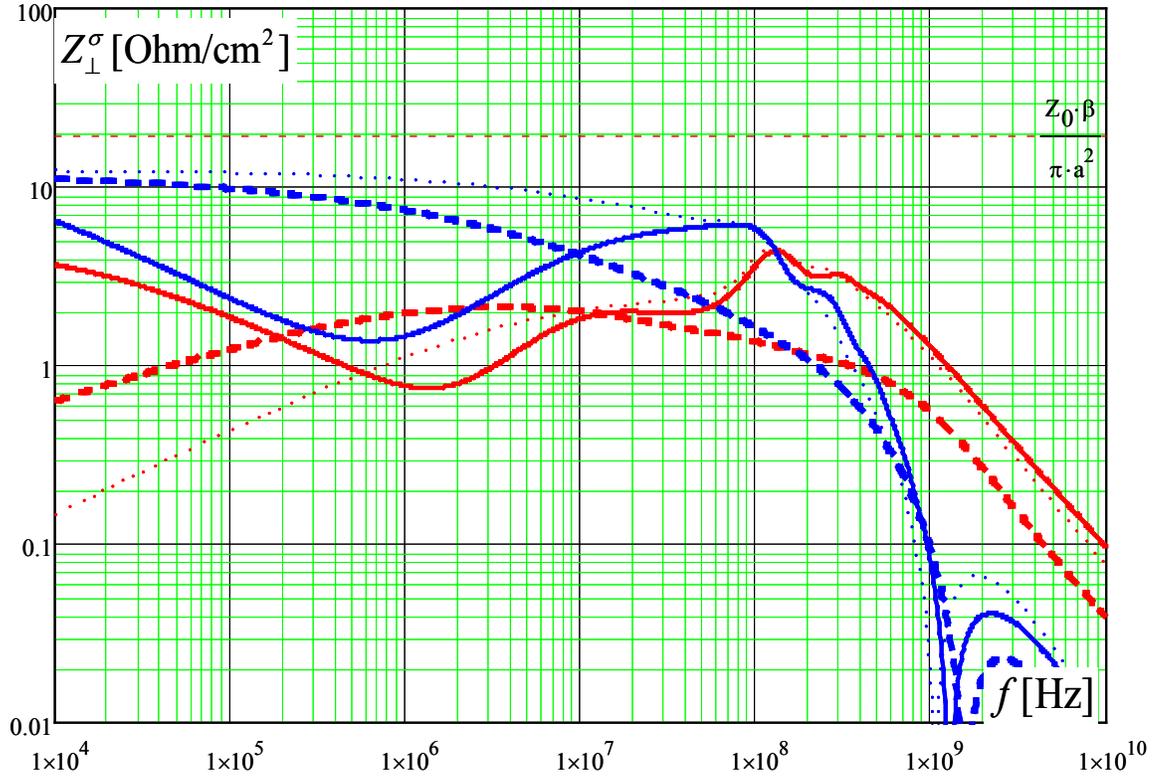

Figure 3: Transverse impedances ($\gamma \to \infty$) for the round (solid lines) and flat geometry (dash lines for the horizontal and dot lines for the vertical). Red lines are for the real parts, blue – for the absolute value of the imaginary parts.

The second reason is change of the field structure at $|Rc/(\omega a)| \sim 1$, equivalent to $\mu\delta/d \sim 1$ or $f \sim 10$ MHz. At low frequencies, when $\mu\delta/d \gg 1$, the fields inside the free space, $r < a$, are of the magnetic type: the magnetic field is almost orthogonal to the magnet surface, $|H_\varphi/H_r|_{r=a-0} \ll 1$. In the opposite case, for $\mu\delta/d \ll 1$, the fields are of the conductivity type: $|H_r/H_\varphi|_{r=a-0} \ll 1$. Interplay of these and some geometrical factors leads to variety of possibilities for impedance behavior at low frequencies seen in Fig. 3. Note, contrary to the longitudinal impedance, the transverse one never exceeds its careless limit $Z_0\beta/(\pi a^2)$. That is why a popular Panofsky-Wenzel estimation of the transverse impedance from the longitudinal is inapplicable here: its use at low frequencies may result in order(s) of magnitude overestimation for the transverse impedance.